\documentclass[lettersize,journal]{IEEEtran}
\usepackage{amsmath,amsfonts}
\usepackage{algorithmic}
\usepackage{algorithm}
\usepackage{array}
\usepackage{textcomp}
\usepackage{stfloats}
\usepackage{url}
\usepackage{verbatim}
\usepackage{graphicx}
\usepackage{subfigure}
\usepackage{cite}  
\usepackage[numbers,sort&compress]{natbib}
\usepackage{amsmath}
\usepackage{cuted}
\usepackage{lipsum} 
\usepackage{color}
\usepackage[absolute,overlay]{textpos} 
\newcommand{\triangleq}{\stackrel{\triangle}{=}}
\usepackage[margin=0.5in]{geometry}
\usepackage{caption} 
\usepackage{times}
\usepackage{hyperref}
\usepackage{titlesec}
\titlespacing*{\subsection}{0pt}{*1}{*0.5} 
\usepackage{indentfirst}
\usepackage{booktabs}
\usepackage[x11names]{xcolor}
\usepackage{comment}
\usepackage{colortbl} 
\usepackage{xcolor}
\usepackage{xpatch}
\makeatletter
\def\changeBibColor#1{%
	\in@{#1}{10838324,10771629}
	\ifin@\color{black}\else\normalcolor\fi
}
\xpatchcmd\@bibitem
{\item}
{\changeBibColor{#1}\item}
{}{}

\xpatchcmd\@lbibitem
{\item}
{\changeBibColor{#2}\item}
{}{}

\makeatother

\begin{document}
\title{\parbox{\linewidth}{\centering Beamforming Design for ISAC Systems with Suppressed Range-Angle Sidelobes}}

\author{{Meihui Liu},~\IEEEmembership{Graduate Student Member, IEEE}, Shu Sun,~\IEEEmembership{Member, IEEE},\\  {Ruifeng Gao},~\IEEEmembership{{Member, IEEE}}, and Meixia Tao,~\IEEEmembership{Fellow, IEEE}
\thanks{M. Liu, S. Sun, and M. Tao are with the Department of Electronic Engineering and the Cooperative Medianet Innovation Center, Shanghai Jiao Tong University, Shanghai 200240, China (e-mail: \{meihui\_liu, shusun, mxtao\}@sjtu.edu.cn). (Corresponding authors: Shu Sun and Meixia Tao)}
\thanks{R. Gao is with School of Transportation and Civil Engineering, Nantong University, and is also with Nantong Research Institute for Advanced Communication Technologies, Nantong 226019, China (e-mail: grf@ntu.edu.cn).}
}



\maketitle

\begin{abstract}
Integrated sensing and communication (ISAC) represents a pivotal advancement for future wireless networks. This paper introduces a novel ISAC beamforming method for enhancing sensing performance while preserving communication quality by leveraging the ambiguity function (AF).  We formulate an optimization problem to minimize the integrated sidelobe level ratio (ISLR) of the AF subject to the constraints of transmission power, communication signal-to-interference-plus-noise ratio, and sensing gain. To address the non-convexity of the optimization problem, semidefinite relaxation is adopted. Numerical results show that our method significantly reduces range sidelobes and achieves a lower ISLR in the range-angle domain compared to other approaches.
\end{abstract}

\vspace{-5pt}
\section{Introduction}
\IEEEPARstart{T}{he} next-generation wireless networks, such as \textcolor{black}{beyond} 5G and 6G, are envisioned to offer not only superior wireless connectivity but also high-precision and robust sensing capabilities in order to support many emerging applications such as autonomous driving and extended
reality \cite{ITU}. \textcolor{black}{Integrated sensing and communication (ISAC)} technology has thus emerged as a key research focus \cite{9737357}.

Waveform design is fundamental to ISAC systems. Early research primarily focused on the time-frequency domain, \textcolor{black}{optimizing existing communication and sensing waveforms \cite{9093221}, or leveraging the inherent sensing potential of communication signals \cite{sturm2011waveform} to achieve dual functionality in communication and sensing.}
\textcolor{black}{With advancements in time-frequency domain research, beamforming has emerged as an advanced and effective approach in the spatial domain to realize ISAC waveform design by shaping the transmit signals toward desired directions.  However, most existing studies in both the time-frequency and beamforming-based spatial domains focus on optimizing sensing metrics under the assumption of the target's true location (such as distance, angle, and Doppler). For example, \cite{10628004} optimizes the performance limits of sensing parameter estimation by deriving and minimizing the Cramér-Rao Bound  for the target’s true parameters, while \cite{9724174} enhances the sensing performance of ISAC systems by calculating the target  signal-to-interference-plus-noise ratio (SINR)  at known target locations and maximizing the SINR. In practical sensing scenarios, where the true target parameters are unknown, the system must perform a full-space scan, generating responses at both target and non-target locations. High responses at non-target locations may lead to false alarms or mask weak targets, significantly degrading detection performance. However, existing designs largely overlook the system’s behavior at non-target locations, a critical aspect that remains underexplored.}

\textcolor{black}{The ambiguity function (AF), which measures signal response in the delay-Doppler domain, offers a comprehensive description of signal characteristics, capturing both the main peak at the target location and sidelobes at non-target locations.  Recent literature \cite{10838324} presents a method for designing ISAC waveforms by optimizing the sidelobes of the AF of pulse shaping, aiming to improve the system's sensing performance while satisfying communication constraints. However, this design is limited to the time-frequency domain and does not fully exploit the spatial domain resources.} Few studies have explored the use of spatial degrees of freedom (DoFs) for effective sidelobe suppression. In \cite{10086626}, beampattern matching and minimum weighted beampattern gain maximization designs are utilized to shape the beam pattern toward the target direction, with the potential to suppress sidelobe levels at non-target angles. Nonetheless, these designs leave sidelobes in the range domain unregulated. \textcolor{black}{Recent studies  \cite{10622269}, \cite{10771629}, have proposed  Symbol-Level Beamforming (SLB) methods based on autocorrelation functions or the AF to reduce range-Doppler sidelobes, but these methods sacrifice communication rates due to the constant modulus constraint and exhibits high computational complexity in symbol-intensive scenarios.}


In this paper, we propose a novel beamforming method in ISAC systems using existing communication systems but with low sidelobes in range and angle domains by using the AF as a tool  for sidelobe characterization and suppression. We first derive the range-angle-Doppler AF for an ISAC waveform with given digital beamforming. Then, we optimize the beamforming matrix to minimize the AF sidelobes in range and angle domains while meeting transmit power limitations, SINR constraints, and beamforming gain requirements in the target direction. To address the non-convexity of the design problem, we employ the semidefinite relaxation (SDR) technique for an efficient solution. Numerical simulations demonstrate that the proposed method significantly reduces range sidelobes and outperforms other design methods in suppressing range-angle sidelobes. Unlike SLB, our approach avoids per-symbol optimization, reducing computational complexity \textcolor{black}{in symbol-intensive scenarios} and preventing the communication rate degradation due to constant modulus constraints.

\textit{Notations:} Boldface letters refer to vectors or matrices. For a square matrix \( \mathbf{S} \), \(\text{Tr}(\mathbf{S})\) denotes its trace, while \(\mathbf{S} \succeq 0 \) indicates that \( \mathbf{S} \) is positive semidefinite. For a matrix \( \mathbf{X} \) of arbitrary size, \( \mathrm{rank}(\mathbf{X}) \), \( \mathbf{X}^H \), and \( \mathbf{X}^T \) denote its rank, conjugate transpose, and transpose, respectively. \( \mathbf{I} \)  denotes an identity matrix with appropriate dimensions. \(\mathbb{C}^{x \times y}\) denotes the space of \(x \times y\) complex matrices. \(\mathbb{R}\) denotes the set of real numbers. \(\mathbb{E}(\cdot)\) denotes the statistical expectation. \(\| \mathbf{x} \|\) denotes the Euclidean norm of a complex vector \(\mathbf{x}\). The symbols $\odot$ and $\otimes$ represent the Hadamard  and the Kronecker products, respectively.

\section{System Model}
A monostatic ISAC system is considered in this work, where a multi-antenna base station (BS) equipped with $M$ antennas serves $K$ communication users and detects a target simultaneously. Let
$\mathbf{s}(t){\in\mathbb{C}^{K\times 1}}$ denote the information symbols intended to the $K$ users at time $t$. It is mapped to the $M$ transmit antennas via a beamforming matrix, $\mathbf{W}=[\mathbf{w}_1, \mathbf{w}_2, ... \mathbf{w}_K]{\in\mathbb{C}^{M\times K}}$, resulting in the signal $\tilde{\mathbf{s}}(t)$, which is then transmitted for simultaneous communication and target detection. The signal emitted by the $m$-th antenna  at time $t$ is expressed as 
\begin{equation}\label{eq:1}\tilde{{s}}_m\left(t\right)=\sum_{k=1}^Kw_{m,k}{s}_k\left(t\right). \end{equation}
The sensing target is assumed to be located at a distance $r_0$ from the first element of the ISAC transmit antenna array, and move with a constant radial velocity $\nu_0$ at an angle $\theta_0$.

\subsection{Communication Model}
It is assumed that each user is equipped with a single antenna. The received signal for the user $k$ at time $t$ is given by
\begin{equation}\begin{aligned}
\label{eq:SINR}
y_k\left(t\right)=\mathbf{h}_{k}^H\mathbf{w}_{k}s_{k}(t)+\sum_{\textcolor{black}{n}\neq k}\mathbf{h}_{k}^H \mathbf{w}_{\textcolor{black}{n}}s_{\textcolor{black}{n}}(t)+{z}_{k}(t),
\end{aligned}\end{equation}
where ${z}_k(t)$ represents additive white Gaussian noise with zero mean and variance $\sigma_k^2$ at time $t$, while $\mathbf{h}_k {\in\mathbb{C}^{M\times 1}}$ denotes the channel vector corresponding to user $k$. The extended Saleh-Valenzuela model is adopted, expressed as
\begin{equation}\textcolor{black}{\mathbf{h}_k=\sqrt{g_k}\sum_{l=1}^{L}\beta_{l,k}\mathbf{a}\left(\phi_{l,k}^\mathrm{tx}\right),}\end{equation}
where $g_k$ denotes the path loss, $L$ represents the number of paths between user $k$ and the BS, $\beta_{l,k}\sim CN(0,\sigma_{l,k}^2)$ represents the complex gain of the $l$-th path, and $\mathbf{a}(\phi_{l,k}^\mathrm{tx}) {\in\mathbb{C}^{M\times 1}}$ is the transmit steering vector with $\phi_{l,k}^\mathrm{tx}$ being the direction of departure (DoD) of the $l$-th path at the BS. It is assumed that the first path is the line-of-sight (LoS) path, while the remaining paths are non-LoS, with the corresponding DoDs $\phi_{l,k}^\mathrm{tx}\sim U(-\pi/2,\pi/2)$, $l=2,..,L$. Treating interference in (\ref{eq:SINR}) as noise, the SINR of user $k$ can be written as
\begin{equation}\label{eq:comSINR}\mathrm{SINR}^{(c)}_k=\frac{|\mathbf{h}_{k}^H\mathbf{w}_k|^2}{\sum_{{\textcolor{black}{n}}\neq k}|\mathbf{h}_{k}^H\mathbf{w}_{\textcolor{black}{n}}|^2+\sigma_k^2}.\end{equation}
The communication $\mathrm{SINR}$ serves as a key metric for evaluating and optimizing communication performance.

\subsection{Sensing Model}
During communication between the BS and the users, signals propagate through free space, reflect off the sensing target, and return to the BS receiver for the sensing task. The impulse response of the sensing channel is given by the following expression, where the delay $\delta(t,\boldsymbol{\Theta}_0)$ is approximated using a Taylor series expansion \cite{san2007mimo}. 
\begin{equation}h\left(t-\delta(t,\boldsymbol{\Theta}_0)\right)\approx h\left(t-\tau_{m,\textcolor{black}{n}}(r_0,\theta_0)+\frac{f_{\nu_0}}{f_c}(t-\tau_{m,\textcolor{black}{n}}(r_0,\theta_0))\right),\end{equation}
where $\mathbf{\Theta}_0=(r_0,\nu_0,\theta_0)$ represents the set of parameters for target distance, velocity, and angle, $f_{\nu_0}=\frac{2\nu_0}{\lambda}$ denotes the Doppler frequency, $f_c$ denotes the carrier frequency, $\lambda$ is the wavelength, and $\tau_{m,n}$ represents the path delay for the $(m,n)$th transmit-receive channel.

Consequently, the received signal by the $n$-th antenna prior to demodulation can be expressed as follows
\begin{equation}\begin{aligned}\label{eq:9}
\tilde{r}_{\textcolor{black}{n}}\left(t,\mathbf{\Theta}_0\right)
=&\sum_{m=1}^{M}\alpha_{m,\textcolor{black}{n}}\tilde{s}_{m}\left({\textcolor{black}{\Big(}}1+\frac{f_{\nu_0}}{f_c}{\textcolor{black}{\Big)}}(t-\tau_{m,\textcolor{black}{n}}\left(r_0,\theta_0\right)\right) \\
&\times\exp\Big\{-j2\pi \tau_{m,\textcolor{black}{n}}(r_0,\theta_0)(f_c+f_{\nu_0})\Big\}\\
&\times \exp\Big\{ j2\pi (f_c+f_{\nu_0})t  \Big\}+\tilde{z}_{\textcolor{black}{n}}(t),
\end{aligned}\end{equation}
where $\alpha_{m,\textcolor{black}{n}}$ is the complex reflection coefficient for the \textcolor{black}{$(m,n)$th}
transmit-receive channel. \textcolor{black}{Under the far-field assumption \cite{Sun},  the path delay can be approximated as follows} 
\begin{equation} \label{eq:10}{\tau}_{m,\textcolor{black}{n}}\left(r_0,\theta_0\right)\approx\tau\left(r_0\right)+\tau_{{\mathrm{T},m}}\left(\theta_0\right)+\tau_{{\mathrm{R},\textcolor{black}{n}}}\left(\theta_0\right),\end{equation}
where $\tau(r_0)=2r_0/c$, $\tau_{\mathrm{T},m}\left(\theta_0\right)=md\sin\theta_0/c$, and $\tau_{\mathrm{R},n}\left(\theta_0\right)=nd\sin\theta_0/c$ are the delays induced by the round-trip distance, the spacing between transmit elements, and the spacing between receive elements, respectively, with $c$ denoting the speed of light.
\vspace{-6mm}
\section{Ambiguity Function}
In this section, we derive the three-dimensional AF for an MIMO ISAC system. Consider a scenario involving a slowly-moving or stationary target, thus satisfying the narrowband assumption \cite{8378571}
\begin{equation}\label{eq:nbcondition}\frac{2\nu_0B_kT}c\ll1,\end{equation}
where $B_k$ and $T$ denote the bandwidth and the duration of the transmitted signal $s_k(t)$, respectively. At this point, the compressive effect of $f_{\nu_0}/f_c$ on the waveform’s complex envelope can be neglected and its impact is only considered for the carrier phase. Under far-field conditions, $\tau_{\mathrm{T},m}\left(\theta_0\right)\ll\tau(r_0)$ and $\tau_{\mathrm{R},n}\left(\theta_0\right)\ll\tau(r_0)$, where both $\tau_{\mathrm{T},m}\left(\theta_0\right)$ and $\tau_{\mathrm{R},n}\left(\theta_0\right)$ significantly influence the phase of the received signal while having a negligible effect on its envelope. Additionally, under both far-field and narrowband assumptions, we can neglect the effect of target reflection coefficients across different transmit-receive channels, i.e., assume that all $\alpha_{m,n}$ in (\ref{eq:9}) are equal to one \cite{li2015ambiguity}. Thus, (\ref{eq:9}) can be reformulated as (\ref{eq:rj}) after demodulation to the baseband,  where $\mathbf{b}_\mathrm{T}$ denotes the transmit steering vector as shown below with $f=f_c+f_{\nu}$.\vspace{-2.6mm}
\begin{figure*}[!b]
\centering
\vspace{-4.3mm}
\noindent\hrulefill
\vspace{-3pt} 
\begin{equation}\begin{aligned} \label{eq:rj}
\tilde{r}_{\textcolor{black}{n}}\left(t,\mathbf{\Theta_{0}}\right) 
=&\sum_{k=1}^{K}\mathbf{b}_{\mathrm{T}}^{{H}}\left(f_{\nu_{0}},\theta_{0}\right)\mathbf{w}_{k}s_{k}\left(t-\tau(r_{0})\right)\exp\left\{-j2\pi\left(f_{c}+f_{\nu_{0}}\right)\left(\tau(r_{0})+\tau_{\mathrm{R},\textcolor{black}{n}}\left(\theta_{0}\right)\right)\right\}\exp\left\{j2\pi f_{\nu_{0}}t\right\}+\tilde{z}_{\textcolor{black}{n}}\left(t\right)
\end{aligned}\end{equation}
\noindent\hrulefill
\end{figure*}

\begin{equation}\mathbf{b}_\mathrm{T}(f_{\nu},\theta)=\begin{bmatrix}e^{j2\pi f \tau_{\mathrm{T},0}(\theta)},\ldots,e^{j2\pi f \tau_{\mathrm{T},M-1}(\theta)}\end{bmatrix}^{{T}}.\vspace{-0.1mm}\end{equation}
The existing AFs for ISAC waveforms are typically derived from the Woodward's AF, defined as
\begin{equation} \label{eq:af}\chi_{\mathrm{ss}}(\Delta\tau,\Delta f_{\mathrm{d}}) \triangleq \int_{\infty}^{\infty}s(t)s^{*}(t-\Delta\tau)e^{-j2\pi \Delta f_{\mathrm{d}}t}\mathrm{d}t.\end{equation}

The AF in (\ref{eq:af}) only characterizes the range-Doppler resolution capability and does not consider the additional spatial DoFs in MIMO-ISAC systems. In this paper, we derive a range-angle-Doppler AF for an MIMO-ISAC system with digital beamforming \footnote{\textcolor{black}{The AF analysis can be directly extended to systems with hybrid analog-digital beamforming.}} and further design the beamforming matrix based on this AF to achieve a balance between S\&C functionalities.
\vspace{-0.5mm}

In the monostatic ISAC system, where the transmitted signals are known to the receiver, the optimal detector is a matched filter for specific target parameters. By applying matched filtering $\tilde{r}_{n}\Big(t,\mathbf{\Theta}_0\Big)$ to the $k$-th waveform ${s}_{k}\left(t\right)$ with target parameters $\mathbf{\Theta}_1=(r_1,\nu_1,\theta_1)$ and the known gain $\gamma_k=\mathbf{b}_\mathrm{T}^H(f_{\nu_0},\theta_0)\mathbf{w}_k$, the received signal component at the $n$-th  antenna corresponding to the $k$-th transmitted waveform is given by
\begin{equation}\begin{aligned}
\hat{r}_{{nk}}(\mathbf{\Theta}_0,\mathbf{\Theta}_1)&=\int\tilde{r}_{{n}}\left(t,\mathbf{\Theta}_0\right)\gamma_{{k}}^*s_{{k}}^*\left(t,\mathbf{\Theta}_1\right)\mathrm{d}t\\&=\hat{r}_{{nk}}^{\prime}(\mathbf{\Theta}_0,\mathbf{\Theta}_1)+\hat{z}_{{nk}},
\end{aligned}\end{equation}
where $\hat{z}_{{nk}}=\int\tilde{z}_{{nk}}\left(t\right)\gamma_{{{k}}}^{*}s_{k}^{*}\left(t,\mathbf{\Theta}_1\right)\mathrm{d}t$.
\begin{figure*}[!b]
\centering
\vspace{-13pt}
\begin{equation}\begin{aligned} \label{eq:AF}
\chi(\mathbf{\Theta_{_0}},\mathbf{\Theta_{_1}}) =&\sum_{\textcolor{black}{n}=1}^{M}{\sum_{\textcolor{black}{k}=1}^{K}}\hat{r}_{\textcolor{black}{nk}}^{\prime}(\mathbf{\Theta}_0,\mathbf{\Theta}_1) \\
=&\sum_{\textcolor{black}{n}=1}^M\sum_{\textcolor{black}{k}=1}^K \int 
\overbrace{ \sum_{\textcolor{black}{k^\prime}=1}^K \mathbf{b}_{\mathrm{T}}^{{H}}\left(f_{\nu_0},\theta_0\right)\mathbf{w}_{\textcolor{black}{k^\prime}}s_{\textcolor{black}{k^\prime}}\left(t-\tau\left(r_0\right)\right)\mathrm{exp}\left\{-j2\pi\left(f_c+f_{\nu_0}\right)\left(\tau\left(r_0\right)+\tau_{\mathrm{R},{\textcolor{black}{n}}}\left(\theta_0\right)\right)\right\}\mathrm{exp}\left\{j2\pi f_{\nu_0}t\right\} }^{\tilde{r}_{\textcolor{black}{n}}\left(t,\Theta_0\right)} \\
&\times \overbrace{ \mathbf{w}_{\textcolor{black}{k}}^{{H}}\mathbf{b}_\mathrm{T}\left(f_{\nu_1},\theta_1\right)s_{\textcolor{black}{k}}^*\left(t-\tau\left(r_1\right)\right)\mathrm{exp}\left\{j2\pi\left(f_c+f_{\nu_1}\right)\left(\tau\left(r_1\right)+\tau_{\mathrm{R},{\textcolor{black}{n}}}\left(\theta_1\right)\right)\right\}\mathrm{exp}\left\{-j2\pi f_{\nu_1}t\right\} }^{\gamma_{\textcolor{black}{k}}^*s_{\textcolor{black}{k}}^*\left(t,\Theta_1\right)} \mathrm{d}t \\
=&\overbrace{ \mathbf{b}_{\mathrm{R}}^{{H}}\left(f_{\nu_0},\theta_0\right)\mathbf{b}_{\mathrm{R}}\left(f_{\nu_1},\theta_1\right)}^{\mathbf{\xi}(\mathbf{\Theta}_0^{(2,3)} ,\mathbf{\Theta}_1^{(2,3)})}
\mathbf{b}_{\mathrm{T}}^{{H}}\left(f_{\nu_0},\theta_0\right)\mathbf{W}\mathbf{X}(\Delta r,\Delta f_d)
\overbrace{\mathbf{W}^{{H}}\mathbf{b}_{\mathrm{T}}\left(f_{\nu_1},\theta_1\right)}^{\mathbf{\Upsilon}(\mathbf{\Theta}_1^{(2,3)})}
\phi(\nu_0,\nu_1,r_0,r_1)
\end{aligned}\end{equation}
\end{figure*}
Define the AF as the coherent summation of all the noise-free matched filtering output pairs $(n,k)$, $n=1,...,M$ and $k=1,...,K$. Thus, the AF can be mathematically expressed as (\ref{eq:AF}), shown at the bottom of this page, in which $\phi(\nu_0,\nu_1,r_0,r_1)= \exp\left\{-j2\pi\left(f_c+f_{\nu_0}\right)\tau\left(r_0\right)\right\}\exp\left\{-j2\pi\left(f_c+f_{\nu_1}\right)\tau\left(r_1\right)\right\}$, and $\mathbf{b}_{\mathrm{R}}(f_{\nu},\theta)=\begin{bmatrix}e^{j2\pi f\tau_{\mathrm{R},0}(\theta)},\ldots,e^{j2\pi f\tau_{\mathrm{R},M-1}(\theta)}\end{bmatrix}^{{T}}$ represents the receive steering vector. $\mathbf{X}(\Delta r,\Delta f_d){\in\mathbb{C}^{K\times K}}$ represents the AF matrix in the range-Doppler domain, with the diagonal elements corresponding to the auto-AFs of each user's transmitted signal and the off-diagonal elements representing the cross-AFs between different users. The element in the $k$-th row and $i$-th column of  $\mathbf{X}(\Delta r,\Delta f_d)$ is $[\mathbf{X}(\Delta r,\Delta f_d)]_{ki} = \int s_k\left(t-\tau\left(r_0\right)\right)s_i^*\left(t-\tau\left(r_1\right)\right)\exp\left\{j2\pi\left(f_{\nu_0}-f_{\nu_1}\right)t\right\}\mathrm{d}t$. It is clear that by defining $\Delta r= r_1-r_0$ and $\Delta f_d= f_{\nu_1}-f_{\nu_0}$, the diagonal of $\mathbf{X}(\Delta r,\Delta f_d)$ corresponds to the Woodward's AF.  

\begin{figure}[ht]
    \vspace{-15.8pt}
    \subfigure[]{
        \includegraphics[width=0.225\textwidth]{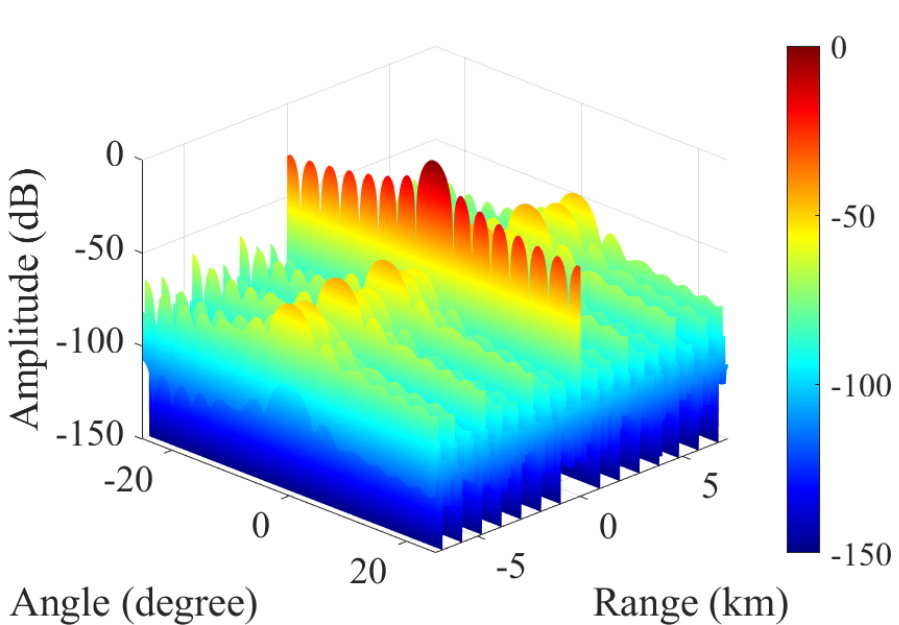}
    }
    \vspace{-10.5pt}
    \subfigure[]{
        \includegraphics[width=0.225\textwidth]{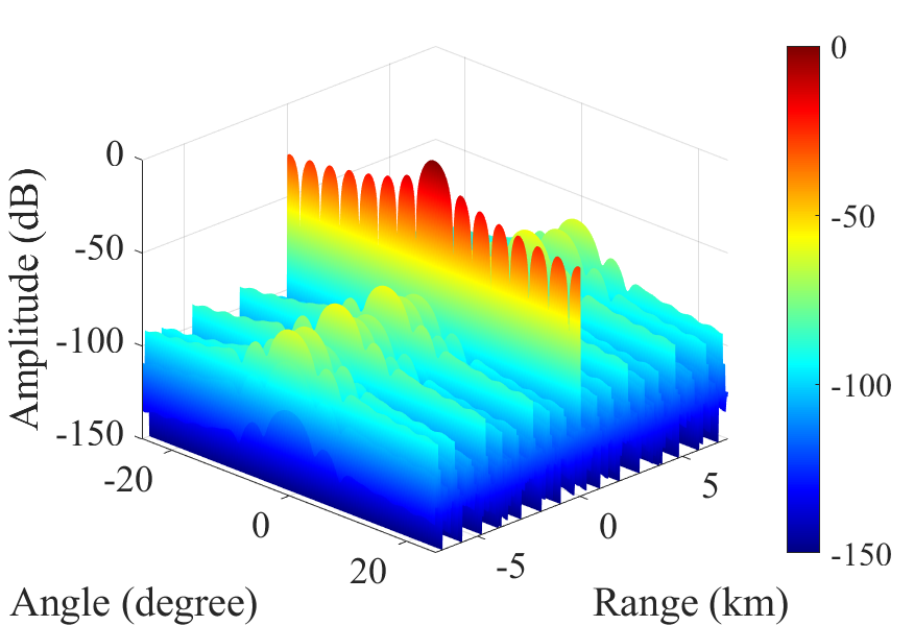}
    }
    \vspace{-8pt}
    \subfigure[]{
        \includegraphics[width=0.225\textwidth]{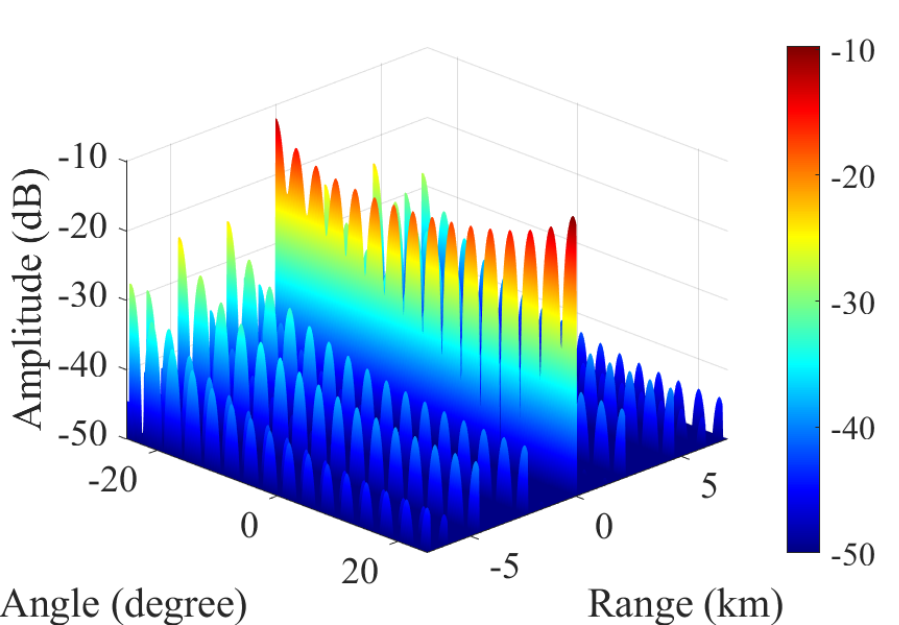}
    }
    \vspace{-2pt}
    \subfigure[]{
        \includegraphics[width=0.225\textwidth]{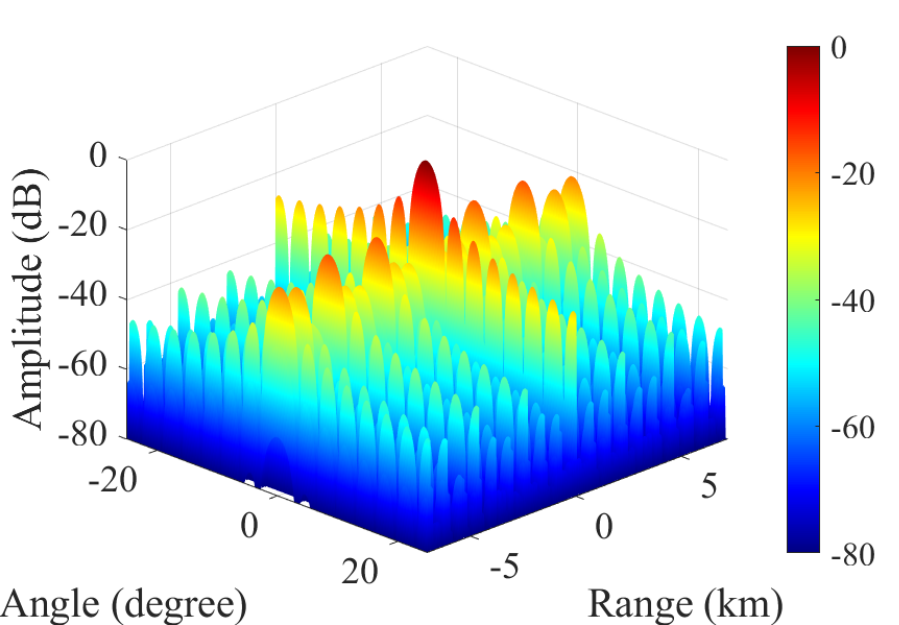}
    }
    \vspace{-7.5pt}
    \subfigure[]{
        \includegraphics[width=0.225\textwidth]{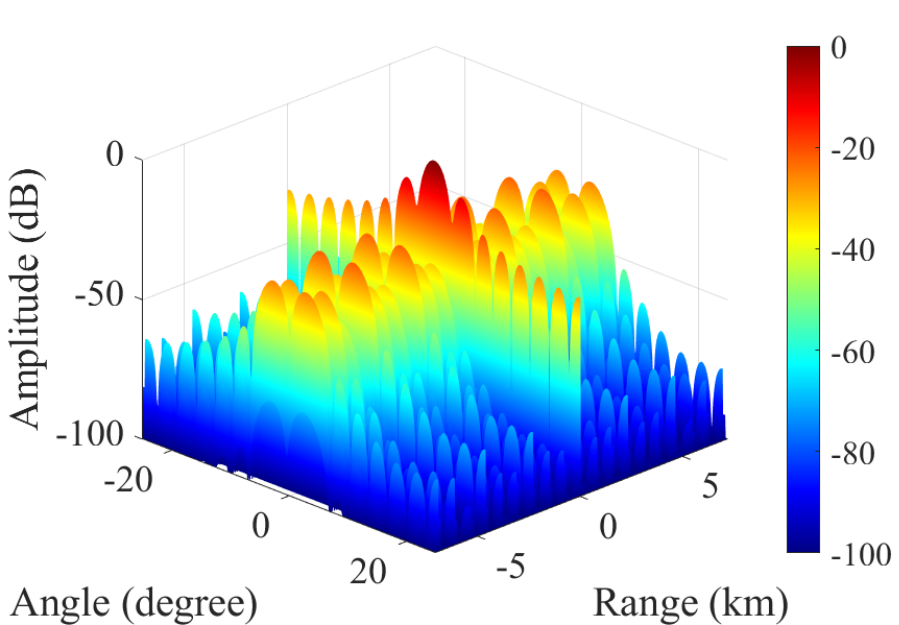}
    }
    \hspace{0.15cm} 
    \subfigure[]{
        \includegraphics[width=0.225\textwidth]{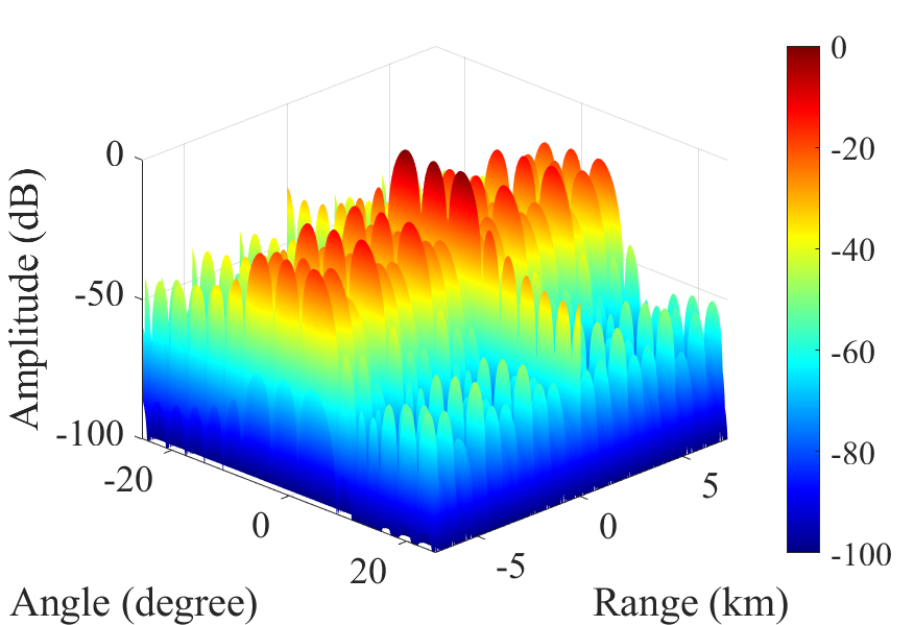}
    }
     \caption{\textcolor{black}{Range-angle AF for different designs. (a) proposed; (b) only sensing: min ISL, s.t. M1 \& M3; (c) only communication: max-min M2, s.t. M1; (d) joint design 1: max M3, s.t. M1 \& M2; (e)-(f) joint design {2, 3}: \{beampattern matching, minimum weighted beam gain maximization\}, s.t. M1 \& M2 in \cite{10086626}. (M1, M2, and M3 correspond to equations (17a), (4), and (16) in this paper, respectively.)}}
     \label{fig:1}
     \vspace{-18.0999pt}
\end{figure}
 \vspace{-3.0999pt}
It is noteworthy that when the transmitted signal $s_k(t)$ is an information-bearing symbol, the AF derived from (\ref{eq:AF}) becomes random. To describe its statistical behavior, one can take the expectation of this function. The resulting expression retains the same form as in (\ref{eq:AF}). Given the similarity between forms $\chi(\mathbf{\Theta}_{_0},\mathbf{\Theta}_{_1})$ and $\mathbb{E}[\chi(\mathbf{\Theta}_{_0},\mathbf{\Theta}_{_1})]$, they are both represented as  $\overline{\chi}(\theta_0,\theta_1,\Delta r,\Delta f_d)$ in this paper. Thus, the AF  is obtained as
\begin{equation}\begin{aligned} 
\overline{\chi}(\theta_0,\theta_1,\Delta r,\Delta f_d) = \xi
{\mathbf{\Upsilon}^H(\nu_0,\theta_0)}
{\bar{\mathbf{X}}(\Delta r,\Delta f_d)}
{\mathbf{\Upsilon}(\nu_1,\theta_1)}
\end{aligned}\end{equation}
where $\xi = \phi
{\mathbf{\xi}(\mathbf{\Theta}_0^{(2,3)} ,\mathbf{\Theta}_1^{(2,3)})}$, and $ \bar{\mathbf{X}}$ represents  $\mathbf{X}$ or  $\mathbb{E}[\mathbf{X}]$. 
To quantify the sidelobe
levels, the \textcolor{black}{integrated sidelobe level (ISL)} is a commonly used metric.   For slowly-moving or stationary scenarios, we focus on the range-angle ISL, denoted as 
\begin{equation}\begin{aligned}\label{eq:ISL0}
\text{ISL} =&\sum_{\theta_1\in\Omega}\sum_{\Delta r\in \mathfrak{D}}\left|\overline{\chi}\left(\theta_0,\theta_1,\Delta r,0\right)\right|^2\
,\end{aligned}\end{equation}
where $\Omega$ and $\mathfrak{D}$ denote the sets of angle sidelobe regions and range sidelobe regions, respectively, to be optimized. 
To achieve accurate sensing, the sensing receiver must receive signals with a sufficient SNR. However, it is challenging to precisely control the SNR at the sensing receiver due to the complexity of the environment. Alternatively, the sensing SNR can be enhanced by controlling the beamforming gain towards the sensing target. The gain in the target direction is given by
\begin{equation}
    P_s = \Big\|\mathbf{b}_\mathrm{T}^{\mathrm{H}}(\nu_0,\theta_0)\mathbf{W}\Big\|^2=\sum_{k=1}^K \Big|\mathbf{b}_\mathrm{T}^{\mathrm{H}}(\nu_0,\theta_0)\mathbf{w}_k\Big|^2.
\end{equation}
\vspace{-10pt}
\section{\textcolor{black}{Beamforming Design for Sidelobes Surppression}}
\vspace{-1pt}
\textcolor{black}{Built on the analysis above, we leverage the spatial DoFs provided by MIMO to design digital beamforming schemes that suppress sensing AF sidelobes and enhance sensing performance.
\vspace{-8pt}
\subsection{Problem Formulation}
We aim to minimize the range-angle ISL of the AF of the transmission signal in ISAC systems, while ensuring that the SINR requirements of communication users, the gain constraints towards the sensing target, and the overall transmission power budget are satisfied. The optimization problem is}
\begin{align}
    (\mathrm{P1}) \underset{\mathbf{\{w\}}_{k=1}^K}{\text{min}} \quad & \mathrm{ISL} \label{op1}\\
    \text{s.t.} \quad
     & \sum_{k=1}^{K}\|\mathbf{w}_{k}\|^2 \leq P_t, \tag{\ref{op1}{a}} \label{op1-a}\\
     & \mathrm{SINR}^{(c)}_k \geq \Gamma_c, k=1,...K,  \tag{\ref{op1}{b}} \label{op1-b}\\
     & P_s = \Gamma_s.  \tag{\ref{op1}{c}} \label{op1-c}
\end{align}
Here $\Gamma_c$ in (\ref{op1-b}) is the minimum SINR constraint for each communication user and $\Gamma_s$ in (\ref{op1-c}) is the beamforming gain constraint for the sensing target \cite{li2015ambiguity},  \cite{7347464} \nocite{10086626}. To facilitate numerical calculations, we discretize $\Delta f_d$ and $\Delta r$ into $L$ points each, transforming  ${\bar{\mathbf{X}}(\Delta r,\Delta f_d)} {\in\mathbb{C}^{K\times K}}$ into the matrix ${\mathbf{Y}}{\in\mathbb{C}^{KL\times KL}}$, where each column represents $L$ range samples at a fixed Doppler frequency.  Meanwhile, $\theta_1$, the angular sidelobe to be optimized, is discretized into $P$ points  within the angular sidelobe region set ${\Omega}=\{\theta_{\mathrm{sl}_1},\theta_{\mathrm{sl}_2},...,\theta_{\mathrm{sl}_P}\}$. As a result, the objective function (\ref{eq:ISL0}) can be reformulated as follows
\begin{equation} \label{ISL_1}
\mathrm{ISL}=\left\|{\mathbf{\Upsilon}^H(\nu_0,\theta_0)}\otimes\mathbf{I}_L\left({\mathbf{Y}}\odot\mathbf{M}\right)(\mathbf{W}^{H}\mathbf{B}_\mathrm{T}\left(f_{\nu_0},{\Omega}\right)\otimes\mathbf{I}_L)\right\|^2,\end{equation}
where $\mathbf M {\in\mathbb{C}^{KL\times KL}}$ is a masking matrix that preserves the sidelobe regions for optimization while nullifying those that do not require optimization. Additionally, $\mathbf{B}_\mathrm{T}\left(f_{\nu_0},{\Omega}\right) = \left[\mathbf{b}_{\mathrm{T}}(f_{\nu_0},\theta_{\mathrm{sl}_1}),...,\mathbf{b}_{\mathrm{T}}(f_{\nu_0},\theta_{\mathrm{sl}_P})\right]$. Under the assumption that the transmission signals are mutually uncorrelated, the cross-AFs of different users are assumed to be nearly zero. Thus,  (\ref{ISL_1}) can be further simplified to  (\ref{eq:ISLsim}), where $[\cdot]_{\{L_1:L_2,L_3:L_4\}}$ denotes the submatrix formed by rows $L_1$ to $L_2$ and columns $L_3$ to $L_4$ of $\cdot$.

\textcolor{black}{Substituting the ISL from (\ref{eq:ISLsim}) into (P1) reveals the non-convexity of the problem, which is inherently challenging due to potential multiple local optima and high computational complexity in achieving global optimality.  However, by applying the SDR technique, the non-convex problem can be reformulated as a convex one. Specifically, we introduce an auxiliary variable $\mathbf{R}_k=\mathbf{w}_k\mathbf{w}_k^{\mathrm{H}},k=1,...,K$ with $\mathbf{R}_k \succeq 0$ and $\mathrm{rank}(\mathbf{R}_k)=1$, the objective function can be reformulated as} 
\begin{figure*}[!b]
\vspace{-5mm}
\noindent\hrulefill
\begin{equation}\begin{aligned}\label{eq:ISLsim}
\mathrm{ISL}
&=\left\|\sum_{k=1}^{K}\mathrm{vec}\Big(\mathbf{b}_{\mathrm{T}}^{{H}}\Big(f_{\nu_{0}},\theta_{0}\Big)\mathbf{w}_{k}\mathbf{w}_{k}^{\mathrm{H}}\mathbf{B}_{\mathrm{T}}\Big(f_{\nu_{0}},{\Omega}\Big)\Big)\Big(\mathrm{vec}\Big([\mathbf{Y}\odot\mathbf{M}]_{\{{(k-1)L+1:kL},{(k-1)L+1:kL}\}}\Big)\Big)^{{H}}\right\|^{2}
\end{aligned}\end{equation}
\end{figure*}
\begin{equation}\begin{aligned}
\mathrm{ISL_{sqt}}=\left\|\sum_{k=1}^{K}\mathrm{vec}\Big(\mathbf{b}_{\mathrm{T}_0}^{{H}}\mathbf{R}_{k}\mathbf{B}_{\mathrm{T_{{\Omega}}}}\Big)\Big(\mathrm{vec}\Big([\mathbf{Y}\odot\mathbf{M}]_{\kappa}\Big)\Big)^{{H}}\right\|,\end{aligned} \end{equation}
where $\mathbf{b}_{\mathrm{T}_0}=\mathbf{b}_{\mathrm{T}}\left(f_{\nu_{0}},\theta_{0}\right)$, $\mathbf{B}_{\mathrm{T_{{\Omega}}}}=\mathbf{B}_{\mathrm{T}}\Big(f_{\nu_{0}},{\Omega}\Big)$ and $\kappa$ denotes region $\{(k-1)L+1:kL,(k-1)L+1:kL\}$. 

Note that (\ref{op1-a}) can be reformulated in terms of $\mathbf{R}_k$ as
\begin{equation} \label{op11}
    \sum_{k=1}^K \mathrm{Tr}(\mathbf{R}_k) \leq P_t.
\end{equation}
Similarly, (\ref{op1-b}) can be expressed as
\begin{equation} \label{op21}
    (1+\Gamma_c^{-1})\mathbf{h}_k^{\mathrm{H}}\mathbf{R}_k\mathbf{h}_k \geq \mathbf{h}_k^{\mathrm{H}}\mathbf{R}_{\mathbf{W}}\mathbf{h}_k + \sigma^2,k=1,...,K.
    \vspace{-1.2mm}
\end{equation}

\begin{table}[h!]
\centering
\caption*{\footnotesize \textcolor{black}{\text{TABLE I}\\SIMULATION PARAMETER CONFIGURATION}} 
\label{tab:1}
\arrayrulecolor{black}
\begin{tabular}{@{}cc@{}}
\toprule
\textbf{\footnotesize \textcolor{black}{Parameter}}                    & \textbf{\footnotesize \textcolor{black}{Value}}              \\ \midrule
\textcolor{black}{Number of antennas, \( M \)}           & \textcolor{black}{36}                          \\
\textcolor{black}{Number of users, \( K \)}              & \textcolor{black}{3}                           \\
\textcolor{black}{Users' LoS DoDs}                           & \textcolor{black}{\( -30^\circ, 30^\circ, 45^\circ \)} \\
\textcolor{black}{Transmit power, \( P_t \)}             & \textcolor{black}{0 dBm}                       \\
\textcolor{black}{SINR for communication, \( \Gamma_c \)} & \textcolor{black}{16 dB }                    \\
\textcolor{black}{Noise power, \( \sigma^2 \) }            & \textcolor{black}{-30 dBm}                     \\
\textcolor{black}{Gain in target direction, \( \Gamma_s \)} & \textcolor{black}{13 dB } \\
\textcolor{black}{Angular optimization region, \( \Omega \)} & \textcolor{black}{\( [-10^\circ, -5^\circ] \cup [5^\circ, 10^\circ] \)} \\
\textcolor{black}{Angular sampling interval}    & \textcolor{black}{0.1°}                  \\
\textcolor{black}{Range optimization region, \( \mathfrak{D} \)}    &\textcolor{black}{\( [-1590 \, \mathrm{m}, -90 \, \mathrm{m}] \cup [90 \, \mathrm{m}, 1590 \, \mathrm{m}] \)} \\
\textcolor{black}{Range sampling rate}          & \textcolor{black}{10 MHz}                      \\
\textcolor{black}{Number of paths per user, \( L \)}                & \textcolor{black}{3}                           \\
\textcolor{black}{LoS path contribution }                  & \textcolor{black}{90\% of total received signal energy }\\ \bottomrule
\end{tabular}
\vspace{-27pt}
\end{table}

If ${\mathbf{w}_k}$ in (\ref{op1-c}) is a feasible solution
to problem (P1), then any phase rotation of ${\mathbf{w}_k}$ is also feasible. Without loss of generality, we choose ${\mathbf{w}_k}$ such that $\mathbf{b}_{\mathrm{T}}^{\mathrm{H}}(f_{\nu_0},\theta_0)\mathbf{w}_k$ is non-negative 
for any user $k$. Consequently, the gain constraint in the direction of the sensing target is given by
\begin{equation} \label{op31}
    \mathbf{b}_{\mathrm{T}}^{\mathrm{H}}(f_{\nu_0},\theta_0)\mathbf{R}_{\mathbf{W}}\mathbf{b}_{\mathrm{T}}(f_{\nu_0},\theta_0) = \Gamma_s.
\end{equation}

\textcolor{black}{With the above re-formulation and relaxing the rank-one constraint on $\mathbf{R}_k$,  the original non-convex problem (P1) is readily transformed into the following convex quadratically constrained semidefinite programming (QSDP) problem (P2),}
\begin{align}
    (\mathrm{P2}) \underset{\{{\mathbf{R}_k}\}_{k=1}^K}{\text{min}} \quad & \mathrm{ISL_{sqt}} \label{op2}\\
    \text{s.t.} \quad 
    &(\ref{op11}), (\ref{op21}), (\ref{op31}) {\tag{\ref{op2}{a}}},\\
    &\mathbf{R}_k \succeq 0, k=1,...K {\tag{\ref{op2}{b}}},\\
    &\mathbf{R}_{\mathbf{W}} = \sum_{k=1}^{K}\mathbf{R}_k. {\tag{\ref{op2}{c}}}
\end{align}

\textcolor{black}{Problem (P2) can be
efficiently solved with guaranteed global optimality using convex
optimization tools like CVX. Notice that the optimal solution of (P2) is not necessarily equivalent to that of (P1). While a rank-one solution of (P2) corresponds to the global optimum of (P1), higher-rank solutions may deviate from (P1)'s solution. Nonetheless, high-quality approximate solutions can still be efficiently obtained through  singular value decomposition or Gaussian randomization.}
\vspace{-11.5pt}
\textcolor{black}{\subsection{Complexity Analysis}}
\vspace{-3.5pt}
\textcolor{black}{The main computational overhead of our method comes from solving the QSDP problem in (P2). Using the primal-dual interior-point method to solve QSDP problem, the worst-case complexity is $\mathcal{O}(K^{6.5}M^{6.5}\log(1/\epsilon))$ for a given solution accuracy $\epsilon$ \cite{9124713}. Compared to the SLB method's complexity of $\mathcal{O}\left\{\ln(1/\epsilon)K^{0.5}N_s^{3.5}M^3\right\}$, where $N_s$ is the number of symbols \cite{10771629}, our approach demonstrates potential advantages in symbol-intensive scenarios but exhibits limitations in large-scale settings due to its higher dependency on the number of users and antennas.}

\vspace{5pt}
\section{Simulation Results}
\vspace{-28pt}
\begin{figure*}[h!]
    \centering
    \begin{minipage}{0.343\textwidth}
        \centering
        \includegraphics[width=\textwidth]{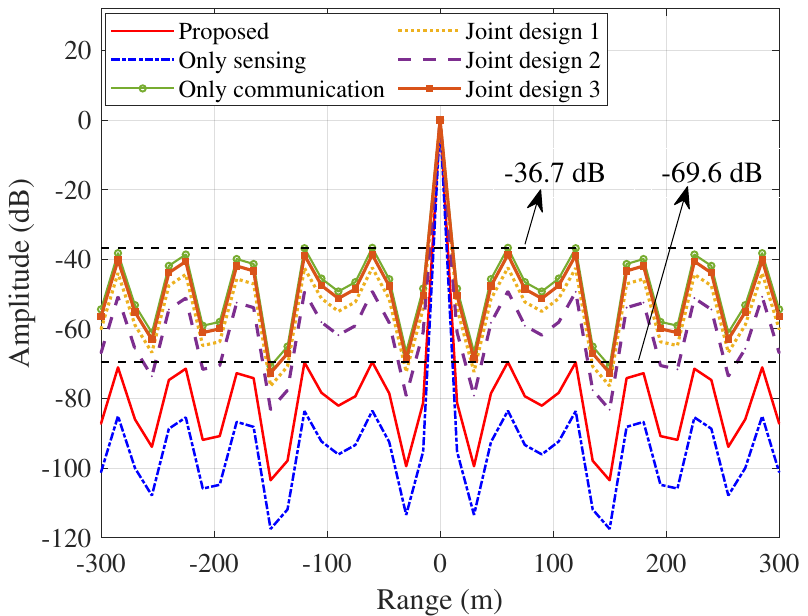}
        \caption{Range cut of the AF for proposed design.}
        \label{fig:2}
    \end{minipage}%
    \begin{minipage}{0.33\textwidth}
        \centering
        \includegraphics[width=\textwidth]{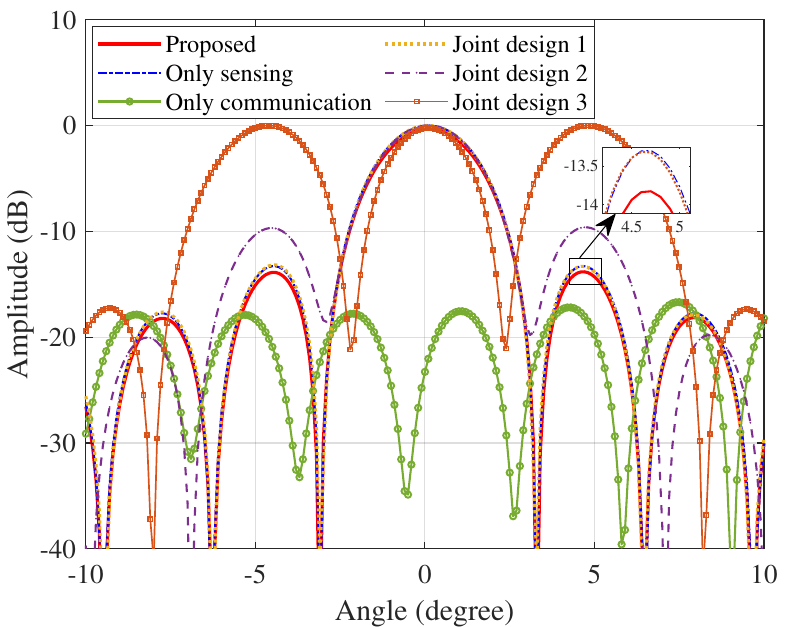}
        \caption{Angle cut of the AF for proposed design.}
        \label{fig:3}
    \end{minipage}%
    \begin{minipage}{0.33\textwidth}
        \centering
        \includegraphics[width=\textwidth]{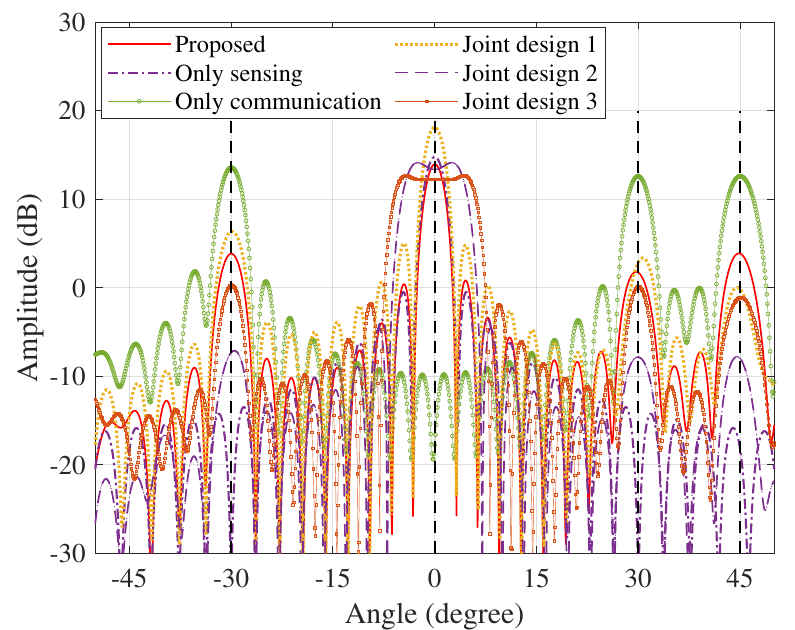}
        \caption{Beampattern  for proposed design.}
        \label{fig:4}
    \end{minipage}
    \vspace{-12pt}
\end{figure*}

\textcolor{black}{This section presents the simulation results obtained using MATLAB, which demonstrate the effectiveness of the proposed beamforming design in suppressing range-angle sidelobes in a multi-user MIMO ISAC scenario. Additionally, the sensing performance of the proposed method is compared with other representative beamforming approaches.} 
\begin{figure}[H]
  \vspace{-4pt}
    \centering
    \includegraphics[width=0.71\linewidth]{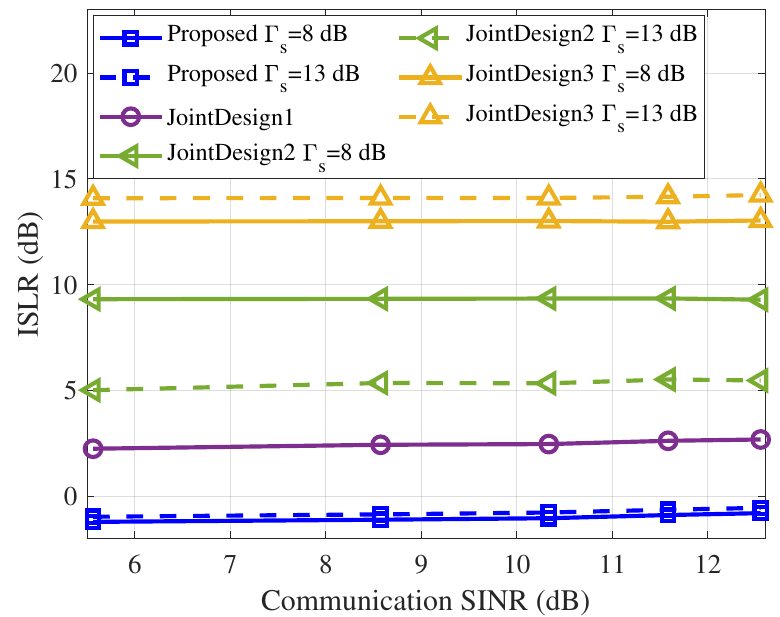}
    \caption{ISL ratio  with communication SINR for different designs}
    \label{fig:5}
    \vspace{-10.5pt}
\end{figure}
In the simulation, the target parameters, $\mathbf{\Theta}_0$ and $\mathbf{\Theta}_1$, share the same Doppler frequency of $0\ \mathrm{kHz}$. \textcolor{black}{For the spatial angles, $\theta_0$ in $\mathbf{\Theta}_0$ is fixed at $0^\circ$, while $\theta_1$ in $\mathbf{\Theta}_1$ varies.  Given the potential of pilot signals in ISAC systems, we employ a Zadoff-Chu sequence of length 512, commonly applied in communication systems. The remaining simulation parameters are shown in TABLE I.}

Fig. \ref{fig:1} depicts the range-angle AFs for the proposed beamforming design and several other beamforming methods.  Fig. \ref{fig:2} and Fig. \ref{fig:3} provide cross-sectional views of the angle and range cuts shown in Fig. \ref{fig:1}. The results indicate that the proposed beamforming design effectively reduces sidelobe levels across the range-angle domain. Notably, it achieves a maximum sidelobe reduction of approximately 33 dB in the range cut compared to designs (c)–(f) and is closest to design (b), which is optimized exclusively for sensing. In the angular cut, the AF of the proposed design significantly reduces sidelobe levels within the optimized angular region compared to designs (e) and (f) and is approximately 0.5 dB lower than that of design (d). Although design (c) achieves the lowest angular sidelobe levels, its extremely low energy near the target results in the poorest sensing performance.

Fig. \ref{fig:4} presents the beampatterns obtained from various design criteria, demonstrating that the proposed method effectively aligns beamforming gains towards both the user and the target. Fig. \ref{fig:5} compares the ISL ratio within the designated optimization region for various design criteria under different sensing gain constraints as the communication SINR varies. The results indicate that for the same communication SINR, the proposed method reduces the ISLR by approximately 3 to 15 dB compared to other methods. Furthermore, as the sensing gain in the target direction increases and the communication SINR improves, the ISLR of the proposed method remains largely unchanged, indicating its capacity to maintain stable sensing performance without significantly compromising communication quality.

\vspace{-4.5pt}
\section{Conclusion}
This paper proposes a novel beamforming method for MIMO-ISAC systems that minimizes the ISLR of the AF, enhancing sensing performance while satisfying constraints on the communication SINR, total transmit power, and sensing gain. By applying SDR to address the non-convex problem, the proposed method significantly reduces range sidelobes and achieves a lower ISLR in the range-angle domain compared to existing counterparts, highlighting its potential for future wireless networks.  
\vspace{-9pt}

\vfill

\begingroup
\footnotesize 
\bibliographystyle{IEEEtran}
\bibliography{ref.bib}

\endgroup

\end{document}